\documentclass[alpha-refs]{wiley-article}

\usepackage{siunitx}
\usepackage{verbatim}

\papertype{Focus Article}

\title{An organized review of key factors for fake news detection}

\author[1\authfn{1}]{Nuno Guimaraes}
\author[1]{Alvaro Figueira PhD}
\author[2]{Luis Torgo PhD}

\affil[1]{CRACS-INESCTEC and University of Porto, Porto, 4200-465 , Portugal}
\affil[2]{Faculty of Computer Science, Dalhousie University, Hallifax, Nova Scotia, NS B3H 1W5, Canada}

\corraddress{Nuno Guimaraes, CRACS-INESCTEC and University of Porto, Porto, 4200-465, Portugal}
\corremail{nuno.r.guimaraes@inesctec.pt}

\presentadd[\authfn{1}]{CRACS-INESCTEC and University of Porto, Porto, 4200-465, Portugal}

\fundinginfo{Nuno Guimaraes thanks the Fundação para a Ciência e Tecnologia (FCT), Portugal for the Ph.D. Grant (SFRH/BD/129708/2017); The work of L. Torgo was undertaken, in part, thanks to funding from the Canada Research Chairs program.}

\runningauthor{Nuno Guimaraes et al.}

\begin{document}

\maketitle

\begin{abstract}
Fake news in social media has quickly become one of the most discussed topics in today's society. With false information proliferating and causing a significant impact in the political, economical, and social domains, research efforts to analyze and automatically identify this type of content have being conducted in the past few years. In this paper, we attempt to summarize the principal findings on the topic of fake news in social media, highlighting the main research path taken and giving a particular focus on the detection of fake news and bot accounts. 

\end{abstract}

\section{Introduction}
The problem of fake news is not recent. In fact, there have been several examples in history before the rise of social media (and the Internet itself). One of the most impactful in modern history was the claim that the HIV virus was fabricated in a United States facility \citep{Boghardt2009}.  This rumor circulated during 1983 and was later captured by a television newscast \footnote{\url{https://www.nytimes.com/2018/11/12/opinion/russia-meddling-disinformation-fake-news-elections.html}}. Although it was posteriorly debunked, the consequences are still present today since some studies suggest the existence of a high percentage of believers in HIV related hoaxes   \citep{hivconspiracy,KLONOFF1999451}. 

With social networks such as Twitter and Facebook, this type of content has platforms where it can be diffused and propagated at a pace that was impossible with other mediums. Furthermore, a recent study concluded that approximately 68\% of American adults use (at least occasionally) social media for their daily news consumption \citep{PewResearch2019}. Consequently, fake news easily reach their target audience and proliferate in this ecosystem, making them one of the most challenging problems in today's society. 

Due to the large quantity of user-generated data in social media, manually verifying all the content published/spread is infeasible. Therefore, researchers are using data mining methods and tools to tackle the fake news problem in this medium.

In this work, we cover the state of the art on social media platforms, to analyse, detect, and minimize the propagation of fake news. We exclude some research topics outside the social media spectrum such as the detection of fake news articles, stance detection, and the development of fact-checking knowledge graphs.

\section{Fake News in Social Media}

Although the first studies on fake news in social media have been published several years before \citep{Castillo,Qazvinian2011}, it was during the 2016 United States presidential election that the term became massively popular. Until then, similar problems were tackled in literature such the analysis and detection of rumors \citep{Fang2013}. However, the fake news concept is slightly different and is more similar to the concept of disinformation (i.e. false information spread or published with the intention of deceiving). Nevertheless, current literature uses the term loosely so, in order to present a more complete review, we include rumor and disinformation as fake news and use the terms interchangeably.

\subsection{Data Annotation}
The growth of fake news led to the rise of fact-checking entities such as Snopes \footnote{https://www.snopes.com} and Politifact \footnote{https://www.politifact.com}, whose purpose is to debunk claims, or Media Bias/Fact Check\footnote{https://mediabiasfactcheck.com/} which offers a large quantity of sources that are known to publish false content. The majority of these fact-checking providers use expert-based annotations in two or more labels. Snopes, for example, has 14 different labels such as True, Mostly True, Mostly False, False, Outdated, Scam, and Unproven. On the other hand, Media Bias Fact Check uses 5 different bias labels (left, left center, least, right center, right) as well as labels like conspiracy, pro-science and questionable sources. 

Several studies in data science use these providers to generate large datasets to study fake news. For example, the dataset LIAR \citep{P17-2067} is composed of 12.8k claims extracted from PolitiFact and the dataset used in \cite{Bovet2019} relies on the annotations of Media Bias Fact Check to extract fake news tweets. Nevertheless, depending on the task, theses sources may not be enough. Thus, several studies rely on experts to manually annotate data. 

It is also worth mentioning that unlike other more traditional text annotation tasks, research in fake news does not commonly uses crowd-sourcing annotation (i.e. rely the annotations to the wisdom of the crowd/volunteers). This can be justified by the complexity of the task and by the fact that unlike other tasks, fake news goal is indeed to deceive the reader, which could lead to poor annotated data if the annotator does not have proper training on the task. 


\subsection{Social Media}
Social media is the main medium for the propagation of fake news and consequently has been a largely studied area. There are several different social media platforms with distinct characteristics and easiness of access to retrieve content. The two most well-known are Facebook and Twitter which are also the main sources of fake news diffusion. The major difference between the two is that Twitter is a microblogging service that allows a 380 characters limit on each post while on Facebook the limit is near 60,000. Also, concerning data accessibility, Twitter has an open API that allows the extraction of data regarding public posts and accounts, something that Facebook has discontinued. This fact, associated with the limitations that the Facebook API imposes (for example, the impossibility to extract posts regarding specific keywords) made that the majority of studies of fake news in social media use Twitter for data extraction.

There is also more region-specific research on Fake News that recur to other social media platforms. For example in Brazil and India, WhatsApp proved to be an important medium on the diffusion of false information \citep{WhatsAppIndia}. This platform is intended to be used as an instant messaging platform. However, in some countries it is more commonly used as a social network since users are added to large groups without knowing all the intervenients. In terms of data accessibility, WhatsApp is a secure and private messaging platform. Therefore, data retrieval must be done manually through automated scripts running on the client. Nevertheless, an analysis of misinformation circulating in WhatsApp groups was conducted by \cite{Resende2019} and even a prototype system for fact-checking has being developed to tackle the problem \citep{Melo2019}. Another example is the Chinese Twitter-like platform Weibo. This social network has similar features to Twitter and allows the use of an API to extract posts. Finally, platforms such as Reddit and 4Chan also proved to be fake news spreaders. One of the most well-known examples was the Pizzagate conspiracy theory that started propagating in these forums \citep{PizzaGateNYT}. Thus, some research in the area also contemplates these platforms as case studies \citep{Kang2015,Dang2016,Zannettou2017}.

In the next section, we elaborate on the different topics inside the area and how the data previously mentioned applies to the research conducted. We will also give particular emphasis to the micro-blogging platforms since they are the focus of the majority of the studies due to the previously mentioned data accessibility issues.

\section{Research Topics}

There are three major areas of research in data science for fake news in social media:
\begin{itemize}
    \item \textbf{the analysis of fake news and accounts that publish fake news} which provides insights about the characteristics of a social media post or account. 
    
     \item \textbf{the detection of false information and accounts whose purpose is to spread false information} normally achieved with the use of machine and deep learning models.
     
     \item \textbf{the analysis of the propagation of fake news throughout the network} with the intention of mitigating them or studying the fact-checking effects in some nodes/users.
\end{itemize}

\subsection{Analysis}
\label{sec:analysis}
The majority of studies that analyse fake news in social media are conducted with respect to a certain event such as the Chile Earthquake \citep{Mendoza2010}, the Mumbai blast in India \citep{Gupta2011}, the bombings at the 2013 Boston marathon \citep{Gupta2013,Starbird2014}, the 2016 United States Election \citep{Bovet2019,zhiwei} and the "Brexit" referendum \citep{Bastos2019,Grcar2017}. The main exception until now is the study published by \cite{Vosoughi2018} which covers a time period from 2006 to 2017.  Several results are important to highlight. First, fake news travel much faster through the network than real or credible news \citep{Vosoughi2018} beginning sometimes with a slow propagation, but once they become viral, their diffusion quickly increases \citep{Gupta2013}. Furthermore, fake news posts tend to increase in important events such as elections \citep{Vosoughi2018,zhiwei}. Concerning fact-checking or credible news diffusion, the majority of the studies agree that fake news propagate faster and in higher quantities than real news or fact-checking content. For example in \cite{Starbird2014}, the authors claim that there is a misinformation to correction ratio of 44:1. This goes against previous findings \citep{Mendoza2010} which support that there is a 1:1 fake news to correction ratio. In a more recent study \citep{Shao2018} the correction is 1:17 thus highlighting an absence of agreement on this subject.  

To better comprehend the users' accountability concerning the problem of 
fake news in social media, we shift our focus to the analysis of accounts that are largely responsible for its dissemination. In \cite{shu2018}, the authors claim that accounts that trust fake news are registered earlier and have high following/followers ratio (i.e. users tend to follow more accounts than to have "followers"). The majority of studies also agree on the importance of social bots for spreading of fake news in social media. Social Bots can be defined as algorithms that produce content automatically and interact with humans. By definition, social bots are not malicious (for example some have the goal of news aggregators). However, malicious social bots have the goal of modifying or influencing behavior, causing a major impact in real-world scenarios whether by shifting public opinion in elections or by affecting the stock market \citep{Ferrara:2016:RSB:2963119.2818717}. It is estimated that the percentage of social bots in Twitter accounts is around 15\% of the users \citep{varol2017online}. Furthermore, in specific events (like elections or tragedies), they act like "super-spreaders" since several studies suggest that a large volume of tweets diffusing fake news can be attributed to a small number of bot accounts \citep{Bastos2019,Shao2017} and the majority of it happens at an early stage (i.e. a few moments after a fake news article is published for the first time). Social bots also present different strategies with respect to the information they spread. A recent study analyzed the key strategies used by social bots to disseminate content on the awakening of an important event (Parkland shooting in Florida). The findings suggest that 36\% of bots retweeted content that criticizes the actors involved in the shooting (such as the police and mainstream media). Other strategies applied by social bots were instilling doubt, sharing reliable information (showing that not all bots are malicious), spreading conspiracy theories, political organization, and commercial gain \citep{Kitzie2018}.   
Nevertheless, it is important to highlight that although social bots amplified discussion on social networks, it is the human-operated accounts that are largely responsible for the diffusion of bot-generated content \citep{Ferrara:2016:RSB:2963119.2818717,Kitzie2018}. On the other hand, since bots are very active sharing fake news at an early stage, a bot classification system capable of a timely detection can be an efficient strategy to avoid the propagation of fake news through the network \citep{Shao2017,Shao2018}. The research towards bots and fake news detection systems is discussed in the next section.

\subsection{Detection}

In terms of detection of fake news in social media we can identify  two main tasks. The first aims to predict if a social network post is false (or a similar concept). More formally given a post with a list of predictors/features $\{X_1,X_2,..X_n\}$ and a target variable $Y$, we aim at approximating the unknown function $f$ such as $Y=f(X_1,X_2,..X_n)$, with $Y$ taking two possible values/labels (i.e. False or True, misinformation/reliable or True/Fake News). In some applications, the target variable $Y$ can have more than two values (fake/reliable/satire) making it a multi-label classification task.

The second task has to do with bots which play an important role on the diffusion of fake news in social media. This detection task has to do with the detection of bots. It is normally approached as a classification task (i.e. label an account as being a bot or human) \citep{davis2016botornot} although there are also studies that approach the problem as a multi-label classification task since they consider an intermediate type of account (cyborg) that is a mainly automated account with rare human intervention \citep{Chu2012}.

\subsubsection{Input Features}
\label{sec:features}
On both tasks, when applying machine learning techniques, it is necessary to analyse and select important features that are able to discriminate among the different class labels. In tasks related to the detection of fake news, we can look at characteristics of the post, the text, and user. Commonly, post-based features include the number of hashtags, mentions, links and weekday of publication. Concerning textual features, besides the number of words and length of the text, sentiment and subjectivity analysis are often used in fake news detection tasks. This is justified by the emotional tone that fake news texts have. Parts-of-speech tags (POS) are also usually extracted, like the number of nouns and pronouns (in 1st, 2nd, and 3rd person) as well as exclamation and question punctuation (due to the absence of formalism in false content). Several studies include a large set of these types of features \citep{Boididou2018,volkova-etal-2017-separating,Knshnan2018,Mendoza2010,Helmstetter2018}. The psychological meaning of words is often analyzed using the LIWC tool \citep{Tausczik2010} due to the psycho-linguistic characteristics of the text. Finally, some studies \citep{Hamidian2015,Helmstetter2018,volkova-etal-2017-separating,zhiwei} also use bag of words or word embedding models to create a large set of features based on the text of the post. 

The third main group of features concerns the user or account that publishes the post as well as the historical behavior of the user. This group is also used in bot detection tasks. Features in this group include the number of followers and friends (since a high number of friends, but a low number of followers can provide cues regarding the type of account), verification status (a verified account is unlikely to be a bot), account age and number of posts (a recent account with a high number of posts could possibly be a bot), \citep{Helmstetter2018,Mendoza2010,Knshnan2018} and the absence/presence of biography, profile picture and banner \citep{Boididou2018}. With data from Weibo, several studies also use the user's gender and  username type \citep{Yang2012,Wu2015,zhang2015}. 
Groups of features used less frequently in fake news detection tasks include propagation and link-based features. Examples include the number of retweets/shares and replies/comments, and the analysis of the cascade of retweets (depth, maximum sub-tree and maximum node) in the social-based group \citep{Mendoza2010}. Link credibility via WOT score \footnote{https://www.mywot.com/}, and Alexa rank \footnote{https://blog.alexa.com/marketing-research/alexa-rank/} is also used \citep{Boididou2018}. 

In bot/fake users detection, beside the user features previously mentioned, additional account-based features are also considered such as the type of client (mobile, web, API...), the number of favorite tweets, and the length of description. In addition, features that analyze default account settings (such as the existence of a profile picture or a banner) and features based on past behavior of users are more frequent. Examples include hashtag, mentions and URL ratios in past tweets \citep{Dickerson2014,shu2018,Chu2012,varol2017online,Ersahin2017,Azab2016}. Features based on the text of past tweets are less common for this task. However, \cite{shu2018} rely on the users' writing style to predict the gender, age, and other psychological characteristics. In addition, the work in \cite{varol2017online} also relies on users past tweets to derive sentiment features for the bot detection task. 

We proceed to summarize the models and evaluation metrics commonly used in the presented detection tasks, as well as the best performances achieved.

\subsubsection{Model Types and Evaluation Metrics}

Fake News Detection is commonly portrayed as a text mining classification task. Therefore, the metrics used for evaluating models built for this task are similar to other text classification tasks, such as sentiment analysis or document classification. True positives, false positives, true negatives, and false negatives are normally computed. However, it is Precision, Recall, F1-score (macro and micro), and Accuracy which are the focus on the evaluation of each system. The use of these metrics are adopted according to the imbalance of used data and the type of task (multi-label or binary). In some studies, the area under the curve (AUC) is also used. 

Models that performed well in more traditional text mining tasks were adopted in the context of fake news detection in social media. For example the studies by \cite{Castillo,Knshnan2018,Hamidian2015} use Decision Trees and achieve an F-measure between 0.83 and 0.86. On the other hand \cite{zhang2015,Wu2015,Yang2012} and again \cite{Knshnan2018}, use Support Vector Machines for the task, accomplishing F1-scores between 0.74 to 0.90. Other approaches include the use of ensemble models (0.9 f1-score) \citep{Helmstetter2018} and Convolutional Neural Networks (0.95 accuracy) \citep{volkova-etal-2017-separating}.
 A more uncommon approach is the harmonic boolean label crowdsourcing presented in \cite{Tacchini2017} that relies on the users' social feedback to predict  if a post is hoax or non-hoax. Although the authors describe excellent results (99\% accuracy), the model presented relies on crowdsourcing the opinion of users based on past behaviour. Thus, it seems unfeasible to apply this model in the absence of social feedback, making it unsuitable in an early detection scenario.

Shifting our focus to the fake users/bot detection task,  this type of task is also generally addressed as a classification task, with several classification algorithms tested and evaluated. In several studies, Random Forests achieve a good performance in distinguishing human and bot accounts with F1-scores ranging from 0.91 to 0.96 \citep{Azab2016,Gilani2017}. Furthermore, the same model proves to be efficient in a three-label classification scenario (human, bot and cyborg) achieving an AUC score of 0.95 \citep{Chu2012}. Naive Bayes is also an often used model accomplishing similar results \citep{Azab2016,Ersahin2017}.

\subsection{Propagation}

 In misinformation or fake news propagation models, the users are commonly illustrated as nodes and the edges the connections of the users to their friends/followers. When a fake news post starts propagating in the network, each node is assigned with a probability of being "affected" by that post. Thus, when analyzing fake news from a network propagation perspective, the problem can be compared with the spreading of an infectious disease where a node (user) can be infected with a certain probability \citep{Kermack1938}. That probability may vary according to several factors. First, not all users believe in "fake news" thus it is important to distinguish them in three classes: the "persuaders" whose goal is to spread and support fake news content, the "gullible users" who are easily influenced by fake news content and "the clarifiers" that are immune to fake news and may confront infected users with fact-checking content 
 \citep{Shu2019}.    
Homophily and social influence theories contribute to the importance of the friends' network in the "fake news contamination" of a gullible user. Accordingly, the probability of a user believing false information can be computed depending on the beliefs of the friends (i.e. a user who has friends that believe in fake news has a higher probability of being infected) \citep{Wu2014book}. Several models and user roles have been proposed based on this approach. \cite{Tambuscio2015} develop a model for the propagation of rumors based on similar user roles (Believer, Fact-checker, Susceptible) and three probabilistic phenomenons: spread (when the user spreads the rumor), verify (when the user fact-checks the rumor), forget (the user forget the news). Another study \citep{Litou2016} considers competing information spreading simultaneous in the network (i.e. the simultaneous spreading of fake news and reliable content). Furthermore, time is an important factor in this model since the probability of a user reading a "fake news" post from its close connections decreases with time.
 
 Some important results arise from these studies. First, fact-checking activity on the network does not need to be in large quantities to cancel the propagation of fake news content and even when a rumor is removed from the network the fact-checking on users that believe the rumor continues \citep{Tambuscio2015}. Second, the percentage of users that are protected against misinformation increases when the propagation time constraints are more relaxed, and it is smaller when the time constraints to spread the information are more restricted  (i.e. when there is an urgency to spread content, more users are infected) \citep{Litou2016}.  These results support and help to explain the results in other fake news analysis studies. Namely that, in the occurrence of an event, the diffusion of fake news tends to occur in higher quantities \citep{Vosoughi2018} and that human accounts are mainly responsible for its propagation \citep{Ferrara:2016:RSB:2963119.2818717,Kitzie2018}.

\section{Conclusion}
\label{sec:disc}

The fake news problem led to an overall increase on the number of studies published in the topic \citep{Figueira2019}. In this work, we present a comprehensive overview of the research and application of data mining techniques for fake news in social media. Although the current studies highlight several important results for understanding fake news, it is our conviction that the problem is still being tackle in a fine-grained fashion and in a time-independent manner, with a focus on event-based analysis and detection. With the exception of the work by \cite{Vosoughi2018}, there seems to be an absence of long term studies around the analysis of fake news in social media. Regarding the models and systems developed, it would be important to evaluate if these can resist the time and change of context in fake news. For example, can models developed in the context of past events be used to tackle the disinformation in social media regarding the 2019 novel coronavirus? We argue that the evolution concerning content and social feedback must be studied to understand if the models and features used, and trained, in past experiences are still applicable today. We do believe that the capability of keeping the relevance of the features and models' performance in different domains and temporal contexts is an essential step towards the detection and mitigation of fake news in social media.

\bibliography{sample}



\end{document}